\documentclass[journal=jctcce,manuscript=article,layout=traditional]{achemso}
\usepackage[version=3]{mhchem} 
\usepackage[table,xcdraw]{xcolor}
\usepackage{fullpage}
\usepackage{mathtools}
\usepackage{hyperref,url}
\usepackage{amsmath}
\usepackage{amsfonts}
\usepackage[capitalise]{cleveref}
\usepackage{amsthm}
\usepackage{amssymb}
\usepackage{bm}
\usepackage{braket}
\usepackage{relsize}
\usepackage{enumitem}
\usepackage[font=small,labelfont=bf]{caption}
\usepackage{graphicx}
\usepackage{footnote}
\usepackage{multirow}
\usepackage{tabularx}
\usepackage{booktabs}
\usepackage{makecell}
\usepackage{threeparttable}
\usepackage{chemfig}
\usepackage{subcaption}
\usepackage{comment}
\usepackage{algpseudocode}
\usepackage{algorithm}
\usepackage{overpic}
\usepackage{svg} 
\usepackage[table]{xcolor} 
\usepackage{upgreek} 
\usepackage{tablefootnote} 
\usepackage{threeparttable}
\def\bv#1{\bm{\mathrm{#1}}} 
\def\asum{\bigoplus_\mathrm{A}^{\mathrm{centers}}}
\def\ov#1{\overline{\bm{\mathrm{#1}}}}
\def\u#1{\underline{\bm{\mathrm{#1}}}}
\def\kdvar#1{\ket{\u{#1}}} 
\def\kuvar#1{\ket{\ov{#1}}} 
\def\buvar#1{\bra{\ov{#1}}} 
\def\oo#1{\overline{\ov{#1}}}
\def\uu#1{\u{\underline{#1}}}

\def\primed#1{{#1}^\prime}
\def\paran#1{\left(#1\right)}

\def\S{\uu{S}}
\def\Sinv{\oo{S}}
\def\Pp{\oo{P}}

\def\Pt{\oo{\boldsymbol{\mathcal{P}}}}

\def\G{\uu{\boldsymbol{\mathcal{S}}}}
\def\Ginv{\oo{\boldsymbol{\mathcal{S}}}}
\def\sap{\bv{S}_\mathrm{AA}^{+1/2}}
\def\sam{\bv{S}_\mathrm{AA}^{-1/2}}
\def\U{\bv{U}_{\mathrm{AA}}}
\def\UU{\bv{U}}

\def\I{\bv{I}}

\def\X{\bv{S}^\mathrm{-1/2}_\mathrm{1C}}
\def\Y{\bv{S}^\mathrm{+1/2}_\mathrm{1C}}

\def\T{\bv{T}}

\def\exp#1{10^{-#1}} 
\def\bond#1#2#3#4{$\text{#1#2#3:}~#4~\text{\AA}$}
\def\ang#1#2#3#4{$\angle\text{#1-#2-#3:}~#4^{\circ}$}
\def\tors#1#2#3#4#5{$\uptau\left(\text{#1-#2-#3-#4:}~#5^\circ\right)$}
\def\CH#1#2{\ce{C_{#1}H_{#2}}}
\def\nane{\CH{n}{2n+2}}

\crefname{equation}{\textbf{Equation}}{\textbf{Equations}}
\crefname{table}{\textbf{Table}}{\textbf{Tables}}
\crefformat{figure}{\textbf{Figure #2#1#3}}

\usepackage{lipsum}
\usepackage{epstopdf}

\author{Anthony O. Lara}
\affiliation{Department of Chemistry, University of California, Berkeley, California 94720, USA}

\author{Justin J. Talbot}
\affiliation{Department of Chemistry, University of California, Berkeley, California 94720, USA}
\alsoaffiliation{Department of Chemistry, Clemson University, Clemson, South Carolina 29634, USA}

\author{Zhe Wang}
\affiliation{Department of Chemistry, University of California, Berkeley, California 94720, USA}
\alsoaffiliation{Chemical Sciences Division, Lawrence Berkeley National Laboratory, Berkeley, California 94720, USA}

\author{Martin Head-Gordon}
\affiliation{Department of Chemistry, University of California, Berkeley, California 94720, USA}
\alsoaffiliation{Chemical Sciences Division, Lawrence Berkeley National Laboratory, Berkeley, California 94720, USA}
\email{mhg@cchem.berkeley.edu}

\title {An algorithm for atom-centered lossy compression of the atomic orbital basis in density functional theory calculations}

\abbreviations{}

\begin{document}

\begin{abstract}
Large atomic-orbital (AO) basis sets of at least triple and preferably quadruple-zeta (QZ) size are required to adequately converge Kohn-Sham density functional theory (DFT) calculations towards the complete basis set limit. However, incrementing the cardinal number by one nearly doubles the AO basis dimension, and the computational cost scales as the cube of the AO dimension, so this is very computationally demanding. In this work, we develop and test a natural atomic orbital (NAO) scheme in which the NAOs are obtained as eigenfunctions of atomic blocks of the density matrix in a one-center orthogonalized representation. The NAO representation enables one-center compression of the AO basis in a manner that is optimal for a given threshold, by discarding NAOs with occupation numbers below that threshold. Extensive tests using the Hartree-Fock functional suggest that a threshold of $10^{-5}$ can yield a compression factor (ratio of AO to compressed NAO dimension) between 2.5 and 4.5 for the QZ pc-3 basis. The errors in relative energies are typically less than 0.1 kcal/mol when the compressed basis is used instead of the uncompressed basis. Between 10 and 100 times smaller errors (i.e., usually less than 0.01 kcal/mol) can be obtained with a threshold $10^{-7}$, while the compression factor is typically between 2 and 2.5.
\end{abstract}

\section{Introduction}
\label{sec:intro}


Kohn-Sham density functional theory (DFT)\cite{Kohn:1996,Capelle:2006,cohen2012challenges} is the leading framework for computational quantum chemistry studies of molecules, as well as condensed matter and interfaces. A DFT model is fully specified by the choice of a functional and a one-particle expansion basis. The most widely used functionals in chemistry are hybrid and range-separated hybrid functionals,\cite{Mardirossian:2017b} which come reasonably close to the so-called chemical accuracy ($\sim 1$ kcal/mol) for reaction energies and barrier heights, while still being quite computationally efficient. Results from benchmark assessments of density functionals against higher-level reference wavefunction theory results, such as the GMTKN55 data set,\cite{Goerigk:2017} the MGCDB84 data set,\cite{Mardirossian:2017b} and the recently reported GSCDB137 data set\cite{liang2025gold} have shown the value of hybrids and range-separated hybrids versus simpler semi-local functionals. 

It is well known that density functionals are developed close to the complete basis set (CBS) limit and typically achieve their best accuracy near it.\cite{Mardirossian:2017b} In the context of atomic orbital (AO) basis sets,\cite{jensen2013atomic,nagy2017basis} the convergence of DFT energies with the cardinal number of the basis is approximately exponential. In practice, a quadruple zeta (QZ) basis is typically required to approach the CBS limit.\cite{Mardirossian:2017b} Larger pentuple zeta (5Z) basis sets generally do not lead to significant changes relative to QZ. As a particular example, the widely used $\omega$B97M-V functional\cite{Mardirossian:2016} was developed using the QZ-level def2-QZVPPD basis set.\cite{Weigend:2005:def2} However, if only a double zeta (DZ) basis is employed, the results using a good hybrid functional, such as $\omega$B97M-V, are significantly degraded relative to QZ, and even a triple zeta basis is not entirely adequate.\cite{Mardirossian:2017b}

Unfortunately, it is still quite common for quantum chemistry calculations using hybrid functionals to be performed with basis sets smaller than quadruple zeta size. The reason is that the computational cost of a calculation on a given molecule increases sharply with the AO basis set size, $N$. Linear algebra steps scale $\mathcal{O}(N^3)$, just like the efficient resolution-of-the-identity (RI) approach to building the Coulomb\cite{eichkorn1995auxiliary} and exact exchange\cite{weigend2002fully,Manzer:2015b} operator matrices. In the AO representation,\cite{weigend2002fully} the latter actually scales as $\mathcal{O}(N^4)$, which is the same as analytical 4-center 2-electron repulsion integral approaches for J and K. Numerical quadrature\cite{becke1988multicenter} to  evaluate the semi-local exchange and correlation contributions scales more favorably, as $\mathcal{O}(N^2)$. Given that increasing the cardinal number by one nearly doubles the size of the AO basis set,\cite{jensen2013atomic} we see that this is associated with roughly an 8 to 16-fold increase in computational cost.

While large basis sets are required to approach the CBS limit, the resulting (enormous) AO density matrices have long been characterized using only minimal basis sets. In particular, it is well recognized that the results of self-consistent field (SCF) calculations can be accurately represented using a molecularly deformed minimum basis.\cite{mcweeny1960some,mulliken1962criteria,davidson1967electronic,roby1974quantum,heinzmann1976population,foster1980natural,ehrhardt1985population,reed1985natural,reed1988intermolecular,mayer1996atomic,mayer1996orthogonal,Maslen1998Locality,cioslowski1998atomic,lee2000extracting,lu2004molecule,laikov2011intrinsic,knizia2013intrinsic,west2013comprehensive,aldossary2022non} Researchers have referred to these representations as natural hybrid AOs,\cite{mcweeny1960some} modified AOs (MAOs),\cite{heinzmann1976population} natural AOs (NAOs),\cite{foster1980natural,reed1985natural,reed1988intermolecular} effective AOs,\cite{mayer1996atomic} polarized AOs,\cite{lee2000extracting} intrinsic AOs (IAOs),\cite{knizia2013intrinsic} etc. The diversity of definitions (and their interesting interconnections\cite{mayer1996atomic,janowski2014near}) shows that a suitable adaptive minimal basis can be derived from an SCF calculation in many ways. The fact that calculations in extended basis sets can be effectively analyzed in terms of minimal basis sets speaks to the high energetic cost of promoting to higher-than-valence principal quantum numbers.


There are two main classes of these molecule-adapted minimal basis sets that are obtained from SCF calculations in large basis sets. The first class imposes the constraint that the molecular minimal basis must exactly span the occupied space, which can be achieved provided that the resulting functions are composed of AOs from \textit{multiple} atoms. That is the case for the IAOs,\cite{knizia2013intrinsic} for example, as well as some other definitions.\cite{lee2000extracting,lu2004molecule,laikov2011intrinsic,west2013comprehensive,aldossary2022non} By contrast, if the minimal basis is constrained to be an \textit{atom-centered} transformation of the underlying AO basis, then the adaptive minimal basis cannot be exact (in general). Many of the earlier proposals\cite{mcweeny1960some,heinzmann1976population,foster1980natural,reed1985natural,ehrhardt1985population,mayer1996atomic} for molecular minimal basis sets fall into this second category.

With this constraint of an atom-blocked transformation from the full AO representation to the molecular minimal basis, either the full AO density matrix cannot be exactly represented, or a variational calculation within the minimal basis will yield a higher energy. Such calculations have been performed,\cite{lee1997polarized,lee2000absolute,berghold2002polarized,kwon2023adaptive} and while far superior to a rigid minimal basis, they cannot be viewed as a fully adequate substitute for a conventional calculation in the native AO basis set. A single-shot correction,\cite{lee2000absolute} using the dual basis approach,\cite{liang2004approaching} can further reduce errors. Surrogate functions\cite{mao2016approaching}, machine learning\cite{schütt2018machine}, and careful parameterization\cite{muller2023atom} have also been applied to replace the (computationally demanding) direct optimization of a small adaptive basis.

In this work, we explore whether on-atom compression of large AO basis sets can be performed without sacrificing significant accuracy in the representation of the density matrix and the associated DFT energy. This will be done by dropping the constraint that the resulting compressed set must be minimal in size. If the result is positive, it would provide a solid foundation for future improvements in the compute efficiency of DFT calculations in large basis sets. We have three design goals in mind. First, the compressed AOs should be atom-centered and thus highly localized, which is beneficial for constructing sparse maps\cite{Pinski:2015,manzer2017general} in low-scaling algorithms. Second, the degree of compression should be controlled by a single threshold, making it easy to also control the error relative to that threshold. Third, the approach should become more effective for large basis sets, aiding the pursuit of the complete basis set limit.

The present goal of on-atom compression of the AO basis for SCF calculations can be contrasted with other widely used approaches to forming compressed representations suitable for efficient computation.  One is to seek an efficient representation for the electron density, $\rho(\mathbf{r})$, where the most prominent example is Coulomb fitting of the density in terms of an auxiliary basis,\cite{baerends1973self,whitten1973coulombic,dunlap1979some,eichkorn1995auxiliary,dunlap2000robust} which has been valuable for accelerating evaluation of the Coulomb energy in DFT calculations.\cite{weigend2006accurate} Alternatives to Coulomb fitting have also been presented.\cite{black2023economical} These methods have the limitations of not being very suitable for exact exchange (which depends on the density matrix), as well as leaving linear algebra steps unaffected. Another alternative is to represent the density matrix in terms of localized molecular orbitals (LMOs), such as by extremizing a localization measure\cite{Edmiston:1963,hoyvik2016characterization} or by Cholesky decomposition of the AO density matrix.\cite{damle2015compressed,fuemmeler2023selected} These methods are valuable in many contexts, including exact exchange evaluation\cite{lin2016adaptively} and post-SCF corrections for double hybrid density functionals.\cite{neugebauer2023assessment} Finally, lossy compression of tensors such as two-electron integrals has been attempted directly in the AO basis,\cite{gok2018pastri} or by approximate factorizations such as tensor hypercontraction.\cite{lee2019systematically,hillers2025lowering,hillers2025accelerating}

The remainder of this paper is organized as follows.  In Sec. \ref{sec:Theory}, we introduce the procedure that performs on-atom compression of an AO density matrix. It is closely related to existing methods that diagonalize atomic blocks of the density matrix in a suitable representation\cite{mcweeny1960some,heinzmann1976population,ehrhardt1985population,mayer1996atomic}. For this reason, we call the compressed functions NAOs, following McWeeny\cite{mcweeny1960some}, although these are \textit{not} exactly the NAOs of natural bond orbital analysis.\cite{foster1980natural,reed1985natural,reed1988intermolecular} From a numerical standpoint, the computational effort necessary to perform this procedure scales linearly with the size of the molecule, so compression will not carry a significant computational burden. The degree of compression obtained is controlled by a single tolerance, $10^{-\epsilon}$, which emerges as the smallest significant NAO occupation number. In Sec. \ref{Sec:Results}, a series of tests is reported which explore the accuracy with which absolute energies and relative energies can be recovered as a function of $\epsilon$ for large AO basis sets ranging from double to triple to quadruple zeta in size. Equally important, the extent of compression for a given $\epsilon$ increases strongly with basis set size, and we characterize the extent of compression possible in the DZ, TZ, and QZ basis sets versus $\epsilon$. Finally, our main conclusions are summarized in Sec. \ref{sec:Conclusions}.

\section{Theory}
\label{sec:Theory}


\subsection{Notation for transformed representations.}
We shall adopt a compact matrix/tensor notation.\cite{head1998tensor} The set of linearly independent AOs can be placed into a row vector whose covariant tensor character is indicated with an underbar, as $\kdvar{\upomega} = \left[\ket{\omega_1}\cdots\ket{\omega_\mu}\cdots\ket{\omega_N}\right]$.  The overlap matrix is likewise covariant and is defined as:
\begin{align}
    \S = \braket{\u\upomega|\u\upomega}
\end{align}
Its inverse has the contravariant (opposite) tensor character (indicated with overbars):
\begin{align}
    \Sinv = \braket{\ov\upomega|\ov\upomega} = \S^{-1}
\end{align}
Evidently, $\ket{\ov\upomega}$ is a row vector of contravariant basis functions defined as a transformation of the given (covariant) AOs:
\begin{align}
    \ket{\ov\upomega} = \kdvar{\upomega} \S^{-1}
\end{align}
The contravariant and covariant basis functions are bi-orthogonal such that $\braket{\omega^i|\omega_j}=\delta^i_{\bullet j}$, or, equivalently:
\begin{align}
    \braket{\u\upomega|\ov\upomega} = \braket{\ov\upomega|\u\upomega} = \I
\end{align}

Finally, it should be noted that quantities without overbars or underbars have neither covariant nor contravariant tensor character and are said to be invariant (e.g., the identity matrix $\I$  in the equation above, and, later, unitary transformations). The action of such quantities does not change the tensor character of the vectors they operate on.



We begin with the AO basis (unprimed quantities, such as $\S$), after removing any near-linear dependencies as described below.  Subsequent transformations will then go to a specially defined one-center orthogonalized AO basis (denoted with primes, such as ${\S}'$). We will then transform to a compressed AO representation (denoted in a different font, such as the compressed overlap matrix, $\uu{\boldsymbol{\mathcal{S}}}$). We describe these transformations below.

\subsection{AO Linear Dependence}
Large basis sets often include a substantial number of diffuse functions.\cite{jensen2013atomic,nagy2017basis,laqua2025conundrum} While these functions are essential for systematically lowering the total energy, their individual contributions are typically small compared to those of more localized basis functions. The inclusion of diffuse functions broadens the eigenvalue spectrum of the overlap matrix, thereby increasing its ill-conditioning. Such near-zero eigenvalues can strongly amplify numerical errors during matrix multiplications, particularly during matrix inversions. 
Below a threshold $\xi$ that is typically $\sim 10^{-8}-10^{-10}$ in double precision arithmetic, it is necessary to remove near-linear dependencies. The most common approach is canonical orthogonalization,\cite{lowdin1970nonorthogonality} but this approach removes linear combinations of basis functions, rather than individual functions. It is therefore incompatible with our atom-centered manipulations.

We thus employ an alternative method, first introduced in the context of obtaining atom-centered localized virtual orbitals\cite{Subotnik:2005virtuals} for use in local correlation methods.\cite{subotnik2006near,wang2023sparsity} This method selectively deletes individual AOs until the smallest eigenvalue of $\S$ exceeds $\xi$, as follows:
\begin{enumerate}
\item While $ \lambda_0 < \xi$.
\begin{enumerate}
    \item Form the overlap matrix $\S$ from the current AO basis, which is the set T.
    \item  Pick the smallest eigenvalue, $\lambda_0$, of $\S$, and its corresponding eigenvector. 
    \item Let $c_0$ be the coefficient of the largest magnitude in this eigenvector. The AO $\omega_0$ with coefficient, $c_0$, contributes most significantly to this most insignificant eigenvector. 
    \item Remove the AO $\omega_0$ from the current AO basis set, T.
\end{enumerate}
\item  T is now a stable subset of AOs. Form the selector matrix $\T$ which selects linearly independent AOs from the complete set.
\item Construct the reduced dimension, numerically stable, overlap matrix: $\S \leftarrow \T^\dagger\S\T$.
\end{enumerate}
All subsequent manipulations are performed in this linearly independent subset of the original AO basis, with its redefined dimension $N \leftarrow N - N_\mathrm{dep}$, where we have removed $N_\mathrm{dep}$ AOs. 

\subsection{One-Center Orthogonalized AO Representations}

To prepare for finding the compressed AOs that best represent the density matrix, we first reduce the representation dependence by performing one-center orthogonalization. First, the AOs are grouped by atomic identity (which can also be generalized to functional groups). The overlap matrix is partitioned into distinct blocks indexed by sites $A, B, \dots$ with respective basis functions $\kdvar{\upomega}_\mathrm{A},\ \kdvar{\upomega}_\mathrm{B},\dots$ where $\kdvar{\upomega}_\mathrm{A}$ each block contains only the subset of AOs associated with $A$, etc. 



Given this one-center partitioned overlap matrix, we can symmetrically orthogonalize within each (diagonal) block to define the one-center orthogonalized AOs:
\begin{align}
     \ket{\primed{\u\upomega}}_\mathrm{A} = \kdvar{\upomega}_\mathrm{A}\sam = \kuvar{\upomega}_\mathrm{A}\sap 
\end{align}
The direct sum of $\sam$ and $\sap$ across all centers yields matrices $\X$ and $\Y$:
\begin{align}
    \X &= \asum\sam \\
    \Y &= \asum\sap
\end{align}
These matrices operate on the contravariant and covariant basis to define one-center orthogonalized representations as:
\begin{align}
    \ket{\primed{\u\upomega}} &= \kdvar{\upomega}\X \\
    \ket{\primed{\ov\upomega}} &= \kuvar{\upomega}\Y
\end{align}
The overlap and its inverse in the one-center orthogonalized representations are:
\begin{align}
    \primed{\S} &= \X\S\X \\
    \primed{\Sinv} &= \Y\Sinv\Y
\end{align}
These matrices are the identity within an atomic block ($\primed{\S}_\mathrm{AA} = \primed{\Sinv}_\mathrm{AA} = \mathbf{\mathrm{I}}_\mathrm{AA}$), but the interatomic blocks are non-zero ($\primed{\S}_\mathrm{AB} \ne \mathbf{\mathrm{0}}_\mathrm{AB}; \ \primed{\Sinv}_\mathrm{AB}\ne \mathbf{\mathrm{0}}_\mathrm{AB}$), reflecting only one-center orthogonalization.

\subsection{Diagonalization of the One-Center Density Matrix}
The conventional AO density matrix is expressed in the contravariant basis as:
\begin{align}
    \Pp &= \buvar{\upomega}\hat{P}\kuvar{\upomega}
\end{align}
In this representation, $\Pp$ is one-centered orthogonalized with $\Y$ to yield:
\begin{align}
    \primed{\Pp} =\Y\Pp\Y = \Y\buvar{\upomega}\hat{P}\kuvar{\upomega}\Y
\end{align}
Within each of its diagonal blocks, $\primed{\Pp}$ is diagonalized to yield atomic occupation numbers, $\bv\uprho_\mathrm{A}$, and associated eigenvectors, $\U$, which can be identified as natural atomic orbitals (NAOs). Specifically:
\begin{align}
    \primed{\Pp}_\mathrm{A} &= \U\bv\uprho_\mathrm{A}\U^\dagger
    \label{eq:NAO_definition}
\end{align}
The union of all the diagonal blocks defines the molecular set of atomic natural orbitals, $\UU$, and their occupation numbers, $\bv\uprho$:
\begin{align}
    \UU &= \asum \U \\
    \bv\uprho &= \asum \bv\uprho_\mathrm{A}
\end{align}
This new basis is defined as:
\begin{align}
    \kdvar{\upchi} &= \kdvar{\upomega}\X\UU
    \label{eq:all_occupations}
\end{align}
With basis functions $\kdvar{\upchi} = \left[\ket{\chi_1}\cdots\ket{\chi_\alpha}\cdots\ket{\chi_N}\right]$.

\subsection{Truncation to Define Compressed NAOs}
The atomic occupation numbers available from diagonalizing the atom-atom blocks of the density matrix in the one-center orthogonalized representation are a direct measure of the importance of each corresponding natural atomic orbital in the density matrix.  We therefore remove eigenvectors associated with eigenvalues $\bv\uprho <10^{-\epsilon}$, where the choice $\epsilon$ will control the fidelity of this truncated, compressed representation of the density matrix. 
Specifically, the use of $10^{-\epsilon} > 0$ enables us to define a truncated basis, but will induce some error in the number of electrons represented by the density matrix in the compressed space, as well as some error in total energies when they are variationally evaluated.

Truncation of the one-center NAO basis yields a smaller basis of size $M$ corresponding to the retained NAOs $\widetilde\UU$ with occupation numbers $\widetilde{\bv{\uprho}} > \exp{\epsilon}$.  The truncated basis is thus simply a subset of $\kdvar{\upchi}$:
\begin{align}
    \ket{\widetilde{\u\chi}} &= \kdvar{\upomega}\X\widetilde{\UU}
\end{align}
Transforming the original $\S$ and $\Pp$ matrices into the new truncated representation yields compressed overlap and density matrices $\G$ and $\Pt$:
\begin{align}
    \G &= \widetilde\UU^\dagger\X\S\X\widetilde\UU \\
    \Pt_\mathrm{trunc} &= \Ginv\widetilde\UU^\dagger\X\S\Pp\S\X\widetilde\UU\Ginv
\end{align}
An SCF calculation in the compressed representation will use the truncated set of one-center NAOs to optimize a compressed DM, $\Pt$. While $\Pt_\mathrm{trunc}$ is not used directly, it can be employed to measure the loss of electrons due to $\epsilon$-based truncation of the complete AO space. This initial loss of electrons relative to the whole space is defined as:
\begin{align}
    \Delta N_e = N_e^\mathrm{trunc} - N_e^\mathrm{full} = \mathrm{Tr}[\Pt_\mathrm{trunc} \G] - N_e^\mathrm{full}
    \label{eq:electron-loss}
\end{align}
It should be reiterated that this loss of electrons is a measure of the error due to DM-based compression of the AO basis \textit{before} use of that basis for an SCF calculation, which yields the correct electron number.

The present method replaces the original AO basis with this transformed, truncated, and linearly independent set obtained via $\bv{X}$. Beyond this basis modification, the SCF procedure, including Fock construction, diagonalization, and density update, proceeds as usual. It is an interesting topic for future work to consider accelerating Fock matrix construction by \textit{directly} using the compressed representation, but in the meantime, it is straightforward to assemble operator matrices in the AO representation, and then transform to the compressed representation to perform updates of the MO coefficients and the density matrix. This can be most easily accomplished by replacing the usual symmetric or canonical orthogonalizer within an SCF code by the $N \times M$ transformation matrix $\bv{X}$ from the AO basis to orthogonalized compressed functions:
\begin{align}
    \kdvar{\widetilde\chi}  =\kdvar{\upomega}\bv{X}= \kdvar{\upomega}\T\X\widetilde\UU\G^{-1/2}
\end{align}




\section{Computational Details}
The compression scheme described above has been implemented both in standalone Python test code and within a development version of the Q-Chem software package.\cite{epifanovsky2021software} For all tests reported here, we employ the following protocol. First, a conventional SCF is performed in the target basis set. Second, 
with a chosen threshold, $\paran{10^{-\epsilon}}$, the converged density matrix is analyzed as described in Sec. \ref{sec:Theory} to define a compressed AO basis. Third, using the compressed AO basis, we perform an additional SCF calculation to obtain the variationally optimal energy in the compressed representation. Fourth, we then assess deviations in absolute and relative energies between the native AO calculation and the compressed AO calculation, as a function of $\epsilon$. Potential compute and storage savings will be determined by the compression factor, $f(\epsilon) = N/M(\epsilon)$.

SCF models are specified by a method and an AO basis. We choose to keep the SCF method fixed as Hartree-Fock. We do not expect appreciable changes with the use of modern density functionals, since all exhibit similar, nearly exponential convergence towards the CBS limit with the highest angular momentum (cardinal number). While any systematic sequence of AO basis sets can be used to explore the extent of compression versus $\epsilon$, we selected Jensen's polarization-consistent (pc) family\cite{jensen2001polarization,jensen2002polarization,jensen2014unifying} for the calculations reported below. A main reason is that the pc-($X-1$) basis sets are available for cardinal numbers $X=1-5$, which is an exceptionally wide range. The quadruple zeta basis pc-3 was used as the default throughout this work, unless otherwise noted. 

Threshold values $\paran{\exp\epsilon}$ from $\exp{4}$ to $\exp{8}$ in increments of $\exp{0.5}$ were scanned to assess the deviations in absolute and relative energies as well as the potential compute and storage savings indicated by $f\paran{\epsilon}$. 
The occupation numbers are to be evaluated on the total density matrix. We report those numbers divided by 2, so that an occupation number of 1 corresponds to a fully occupied level (in the absence of interatomic overlap effects, to be discussed later).

All SCF calculations were performed using a development version of Q-Chem,\cite{epifanovsky2021software} with the following numerical thresholds to control accuracy and stability. 
Shell pair overlaps were neglected below a value of $\exp{16}$, while two-electron integrals $\paran{\mu\nu|\lambda\sigma}$ were screened using an integral-cutoff of $\exp{14}$. Near linear dependencies in the AO basis were identified using an overlap-eigenvalue cutoff of $\exp{6}$ as described in Sec. \ref{sec:Theory}. The SCF calculation was deemed converged when the electronic wave function error fell below $\exp{8}$.

A variety of benchmark systems are used in this work, as summarized in Table \ref{tab:benchmark-descriptions}. The alkanes (\nane), polyene oligomer (\CH{30}{32}), and polyyne oligomer (\CH{30}{2}) are model linear systems to investigate error extensivity. The ACONF20, BRS36, C20C24, INV23, HSG, and H2O20Rel9 sets are used for relative energies.


\begin{table}[h]
    \centering
    \begin{threeparttable}
    \begin{tabular}{l|ll}
         \textbf{Benchmark}&  \textbf{Type}&\textbf{Description}\\
         \hline
         \hline
         \CH{n}{2n+2}\tnote{1,2}& Single Point&A set of $n=1, 2, 4, 10, 30$ linear polyanes\\
         \CH{30}{32}\tnote{3}& Single Point&A 30 carbon polyene oligomer\\
         \CH{30}{2}\tnote{4}& Single Point&A 30 carbon polyyne oligomer\\
         ACONF20\cite{ehlert2022conformational}&  Isomerization&Isomerization energies of \CH{20}{42} alkane conformers\\
         BSR36\cite{Goerigk:2017,liang2025gold}&  Thermochemistry&Hydrocarbon bond separation reaction energies\\
         C20C24\cite{Manna:2016,liang2025gold}&  Isomerization&Isomerization energies of \ce{C_{20}} and \ce{C_{24}} isomers\\
         INV23\cite{Goerigk:2016,liang2025gold}& Barrier Height&Inversion barrier heights\\
         HSG\cite{Marshall:2011,liang2025gold}& Noncovalent&Binding energies of ligands with protein receptors\\
         H2O20Rel9\cite{Lao:2015,liang2025gold}&  Noncovalent&Isomerization energies of 9 \ce{(H_2O)_{20}} structures\\
    \end{tabular}
    \begin{tablenotes}
    \item[1] $\mathbf{CH_{4}}$ \textbf{Geometric Parameters}. \bond{C}{-}{H}{1.10}, \ang{H}{C}{H}{109.5}
    \item[2] $\mathbf{C_{n}H_{2n+1}}$ \textbf{Geometric Parameters.}~\bond{C}{-}{C}{1.54}, \bond{C}{-}{H}{1.10}, \ang{C}{C}{C}{109.5}, \ang{H}{C}{H}{109.5}, \ang{H}{C}{C}{109.5}, \tors{C}{C}{C}{C}{180}
    \item[3] $\mathbf{C_{30}H_{32}}$ \textbf{Geometric Parameters.} ~\bond{C}{-}{C}{1.42}, ~\bond{C}{=}{C}{1.35}, ~\bond{C}{-}{H}{1.10}, ~\ang{H}{C}{H}{123.12}, ~\ang{H}{C}{C}{34.15}, ~\ang{C}{C}{C}{124.5}, ~\tors{C}{C}{C}{C}{180}
    \item[4] $\mathbf{C_{30}H_2}$ \textbf{Geometric Parameters.} ~\bond{C}{-}{C}{1.36}, ~\bond{C}{$\equiv$}{C}{1.20}, ~\bond{C}{-}{H}{1.06}, ~\ang{C}{C}{C}{180}, ~\ang{H}{C}{C}{180}, ~\tors{C}{C}{C}{C}{180}
    \end{tablenotes}
    \end{threeparttable}
    \caption{Summary of benchmark systems used. \nane~and \CH{30}{32} were generated using a QChem utility script with standard geometries.\cite{pople1967molecular} \CH{30}{2} was constructed using standard geometries. The \nane~set, \CH{30}{32}, and \CH{30}{2} are in their rigid non-optimized geometries. Modified versions of the BSR36, C20C24, INV23, HSG, and H2O20Rel9 benchmarks were drawn from a larger benchmark collection, which also provided the stoichiometric coefficients used for the relative-energy calculations.\cite{liang2025gold}}
    \label{tab:benchmark-descriptions} 
\end{table}

\section{Results}
\label{Sec:Results}


\subsection{Characterization for linear hydrocarbon oligomers}

We select hydrocarbon chain molecules as a suitable test system, in which alkanes, polyenes, and polyynes allow us to assess the role of conjugation and band gap on the one hand, and the effect of differing numbers of nearest neighbors for interior carbon atoms on the other. Long-chain molecules should be used for assessment, because compression is artificially effective in small molecules (e.g., in methane, only 5 compressed orbitals are required on C or H because there \textit{are} only 5 MOs). Considering the high computational effort needed for extended basis sets such as pc-4, we choose chains of 30 C atoms for the tests reported below.

\begin{figure}[h]
    \centering
    \includegraphics[width=1\linewidth]{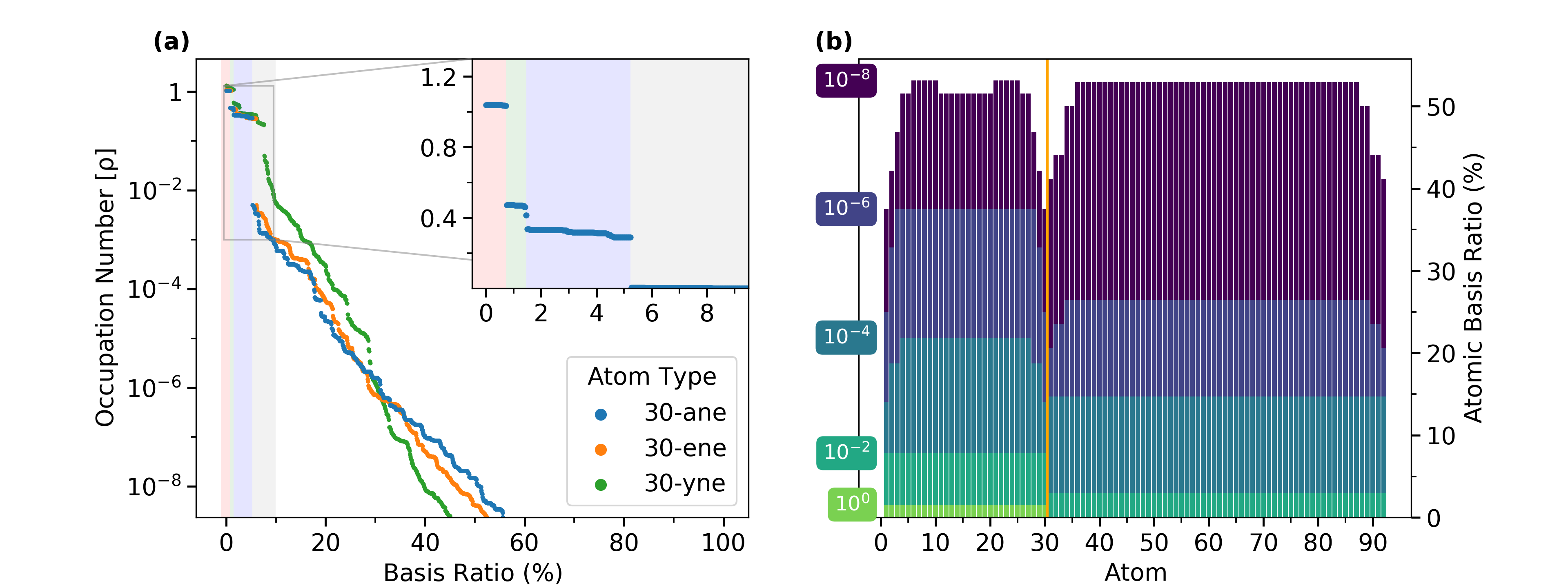}
    \caption{
    Distribution of occupations for the NAOs of 30 chain hydrocarbon oligomers. (a) Occupation number decay for a \CH{30}{62}, \CH{30}{32}, and \CH{30}{2} versus basis ratio (i.e., the x-axis counts eigenvectors in order of occupation number). The inset plot shows a zoomed-in view of the largest occupation numbers for the minimal basis of \ce{C30H62}. (b) Per atom basis ratios of \CH{30}{62} as a function of threshold $\paran{\exp{\epsilon}}$, where the x axis is the atom index, with the 30 C atoms first, followed by the 62 H atoms.
    }
    \label{fig:xnes-dist}
\end{figure}

The first question to consider is the behavior of the NAO occupation numbers obtained from Eqs. \ref{eq:NAO_definition} and \ref{eq:all_occupations}. Figure \ref{fig:xnes-dist}(a) compares the occupation numbers for the \CH{30}{62}, \CH{30}{32}, and \CH{30}{2} molecules (in the pc-3 basis). Their overall behavior is broadly similar, with a small percentage ($\sim 5-8$\%) of large eigenvalues, followed by an extended region of approximately exponential decay in the occupation number with increasing eigenvalue count. This rapid decay in the magnitude of the smaller eigenvalues makes the compression scheme promising in this pc-3 basis. In detail, there are differences.  The \CH{30}{2} exhibits a smaller gap in occupation numbers between the large and small sets than the alkane and polyene. The rate of decay in the small occupation numbers for the polyyne is noticeably quicker as well. 

The inset Figure \ref{fig:xnes-dist}(a) zooms in on the large occupation numbers for the alkane case, and, interestingly, reveals how they originate from different AO types. The fully occupied levels in the left-most red region are C(1s) orbitals (minor deviations from unity are due to effects of inter-atomic overlap). The remaining large occupation numbers derive from the valence C(2s), C(2p), and H orbitals. 
A 1:5 ratio in the number of members of each of the two classes of significantly non-zero eigenvalues suggests the NAOs are AO-like. The smaller set (whose members are equal to the 1s set) is shown in green and is 2s-like. The larger set is shown in blue and divides into 2:2:1 portions whose values are almost identical, corresponding to H(1s), C(2p$_x$), C(2p$_y$), and C(2p$_z$). Together, these largest occupation numbers correspond to the effective minimal basis. The grey area to the right of the blue region corresponds to the most important beyond-minimal functions whose occupations are decaying approximately exponentially with NAO count, as shown in Figure \ref{fig:xnes-dist}(a).

\begin{figure}[h]
    \centering
    \includegraphics[width=0.5\linewidth]{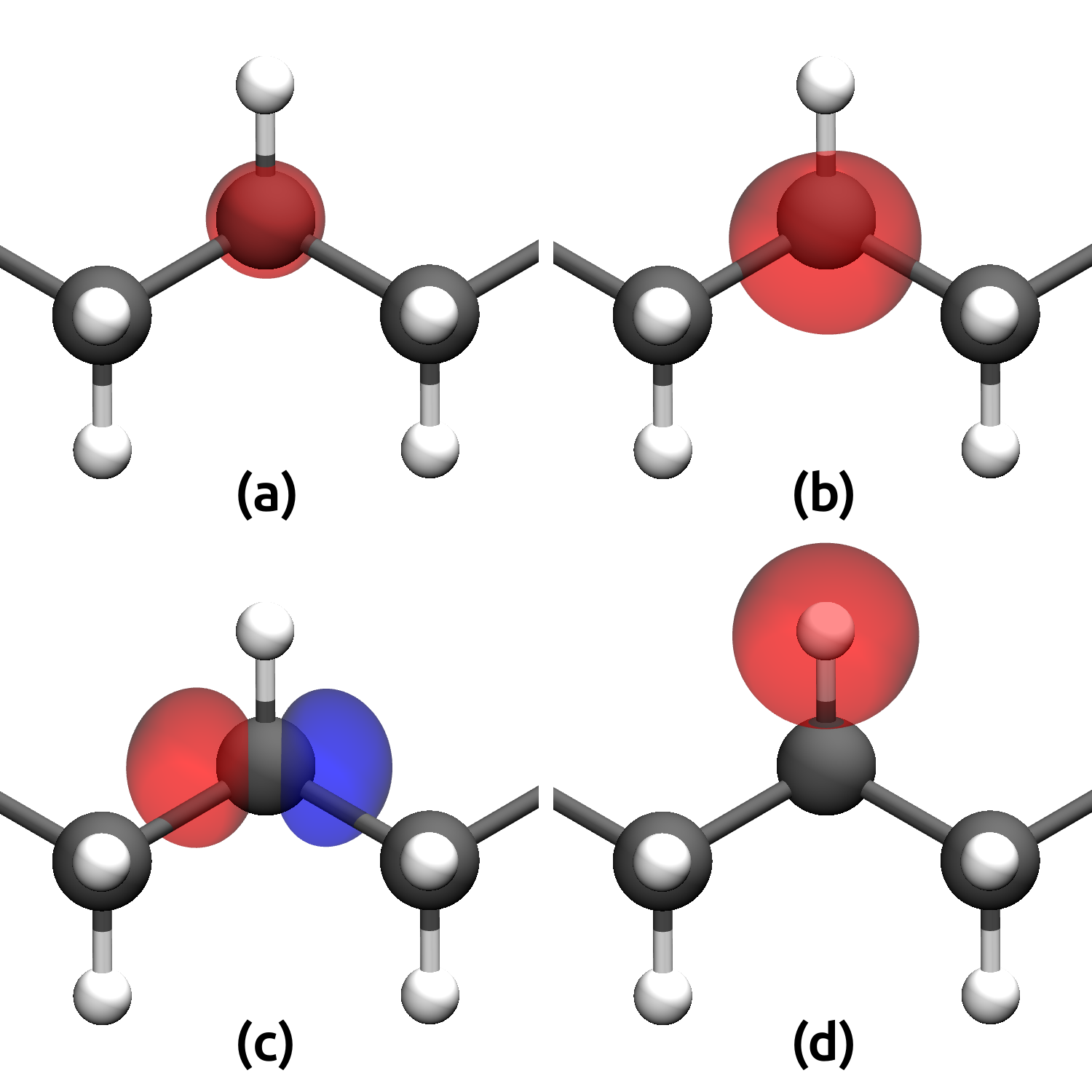}
    \caption{The most strongly occupied NAOs for the innermost carbon $\paran{\mathrm{C15}}$ and its attached hydrogen in \CH{30}{62}. Orbitals (a), (b), and (c) correspond to the three most highly occupied C NAOs, listed in decreasing order of significance. NAOs (a) and (c) retain clear C(1s) and C(2p)-like shapes, while (b) appears as a polarized variant of the C(2s) orbital. For hydrogen (d), the dominant NAO is a slightly distorted version of atomic H(1s) AO. These molecular NAOs thus resemble the corresponding free atom AOs, with small environment-induced perturbations.}
    \label{fig:orbs}
\end{figure}

Visualization of these most important NAOs presented in Figure \ref{fig:orbs} confirms their AO-like nature inferred above. The orbitals clearly resemble distorted versions of the free atom AOs, with characteristic shapes. The perturbations due to forming local bonds and polarizing in the non-spherical environment of the molecule are not enough to fundamentally change the valence atomic orbitals. It is interesting that they are not naturally hybrid orbitals.

Figure \ref{fig:xnes-dist}(b) summarizes how different threshold choices affect the fraction of NAOs retained in the compressed basis for the \CH{30}{62}, where the x-axis is the atom index, with the 30 carbons first, followed by the 62 hydrogens. At the loosest threshold, 1, only the fully occupied C(1s) AOs are retained. At a still-very loose threshold, $\exp{2}$, the truncated basis successfully reconstructs the minimal basis, keeping both the H(1s) and C(2s,2p) valence orbitals discussed above with Figure \ref{fig:xnes-dist}(a). As the thresholds become more stringent, additional NAOs are retained. Interestingly, the retained distribution becomes uneven across the molecule. The fraction of retained NAOs is smaller at the edges for a given threshold. Interior carbons and their attached hydrogens compress less efficiently than terminal groups, reflecting the increased number of ($n^\mathrm{th}$) neighbors. Hydrogen orbitals are consistently more compressible than carbon orbitals at the same threshold levels, reflecting the fact that a given H contributes predominantly to only a single occupied MO.

\begin{figure}[h]
    \centering
    \includegraphics[width=.5\linewidth]{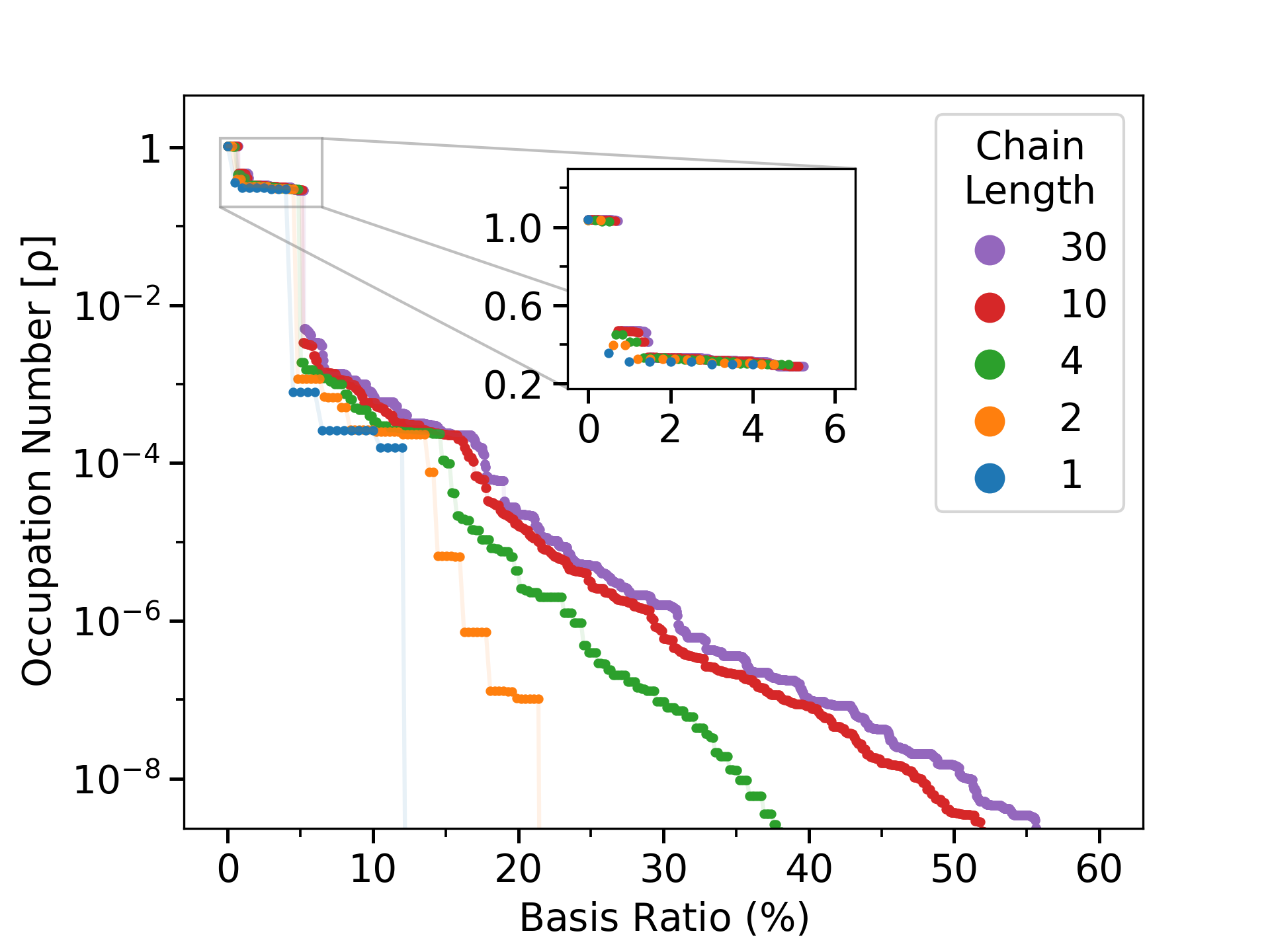}
    \caption{
    Decay of \nane~of increasing chain length as a function of basis ratio.
    Spectral decay for numerically significant occupation numbers. Small chains show remarkable compressibility, which asymptotically decays for the largest chain.
    The inset plot shows a minimal basis of significant occupation is recovered for each \nane. The stratification of the most significant occupation numbers is evident in both the smallest and the largest alkanes.
    Construction of the minimal basis is invariant to system size, while compressibility and corresponding occupation decay rate are dependent on system size. This size dependence asymptotes for moderately sized systems (chain size of about 10). 
    }
    \label{fig:n-anes-decay}
\end{figure}

To further explore how the extent of compressibility depends on the effective number of neighbors, Figure \ref{fig:n-anes-decay} shows how the occupation number distributions depend on increasing carbon chain length. As the chain increases, the decay rate of the occupation spectrum becomes progressively slower, indicating a slower drop-off in orbital significance. This trend continues until approximately ten carbons, beyond which the decay profile begins to plateau. For this 1-dimensional connectivity, this suggests that we have approached a limiting (bulk-like) regime. At the same time, the gap between the minimal basis orbitals and the virtual space decreases slightly with chain elongation. Consistent with our analysis of Figure \ref{fig:xnes-dist}(b), we see that overall compressibility diminishes with increasing system size. Small molecules such as methane are highly compressible, whereas larger chains become increasingly less compressible as the number of distant neighbors with non-negligible interactions increases. 
The inset in Figure \ref{fig:n-anes-decay} compares the composition of the minimal basis as the chain length increases. The carbon 1s occupations are essentially constant with chain length. There is slightly greater variation in the C(2s) occupations with size. Overall, the molecule-adapted minimal basis emerges cleanly and similarly for all chain lengths (note the spacing between points is a consequence of representing the basis count as a percentage, and does not reflect any significant underlying change in occupations).


\begin{figure}
    \centering
    \includegraphics[width=.5\linewidth]{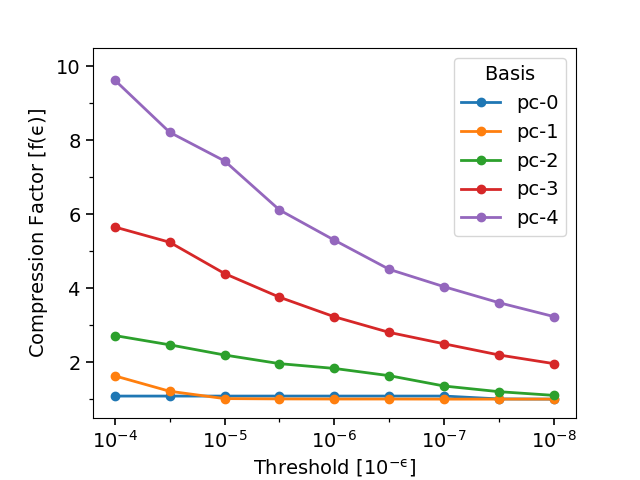}
    \caption{Compression factor as a function of threshold for bases or increasing cardinality for a \CH{30}{62}. The smallest bases, pc-0 and pc-1, exhibit little to no compressibility for our working threshold range. The larger bases, pc-2, pc-3, and pc-4, increase compressibility and yield larger compression factors for our given range. The larger the basis, the greater the fraction of NAOs that are insignificant in describing the SCF energy and density.}
    \label{fig:30-ane-cardinality}
\end{figure}

With compressibility characterized in the large QZ-sized pc-3 basis, we turn next to how compressibility changes with cardinal number in the pc-X sequence. Figure \ref{fig:30-ane-cardinality} examines the compressibility of a 30-carbon alkane as a function of X, via the compression factor $f\paran{\epsilon}=N/M(\epsilon)$, which directly measures how aggressively the basis is reduced at a given threshold, $10^{-\epsilon}$. The smallest bases, pc-0 and pc-1, exhibit little or no compressibility across the entire threshold range, indicating that they lack sufficient redundancy to benefit from the compression procedure. The pc-2 basis shows compression factors as large as 2.5 for relatively loose thresholds such as $\epsilon=4$, suggesting some degree of redundancy in the AO representation that increases smoothly as the threshold is relaxed. 
Compression factors increase sharply between successive basis sizes greater than pc-2 at all threshold values. Notably, pc-4 achieves exceptionally high compression, reaching a factor of over 8 at $\epsilon = 4.5$ (vs over 4 for pc-3) and still yielding a factor of over 4 at $\epsilon = 7$ (vs 2.5 for pc-3). These results demonstrate that higher cardinality bases offer significantly more compressible structure, enabling substantial reductions in basis size without, potentially, severely compromising accuracy.

\begin{figure}
    \centering
    \includegraphics[width=1\linewidth]{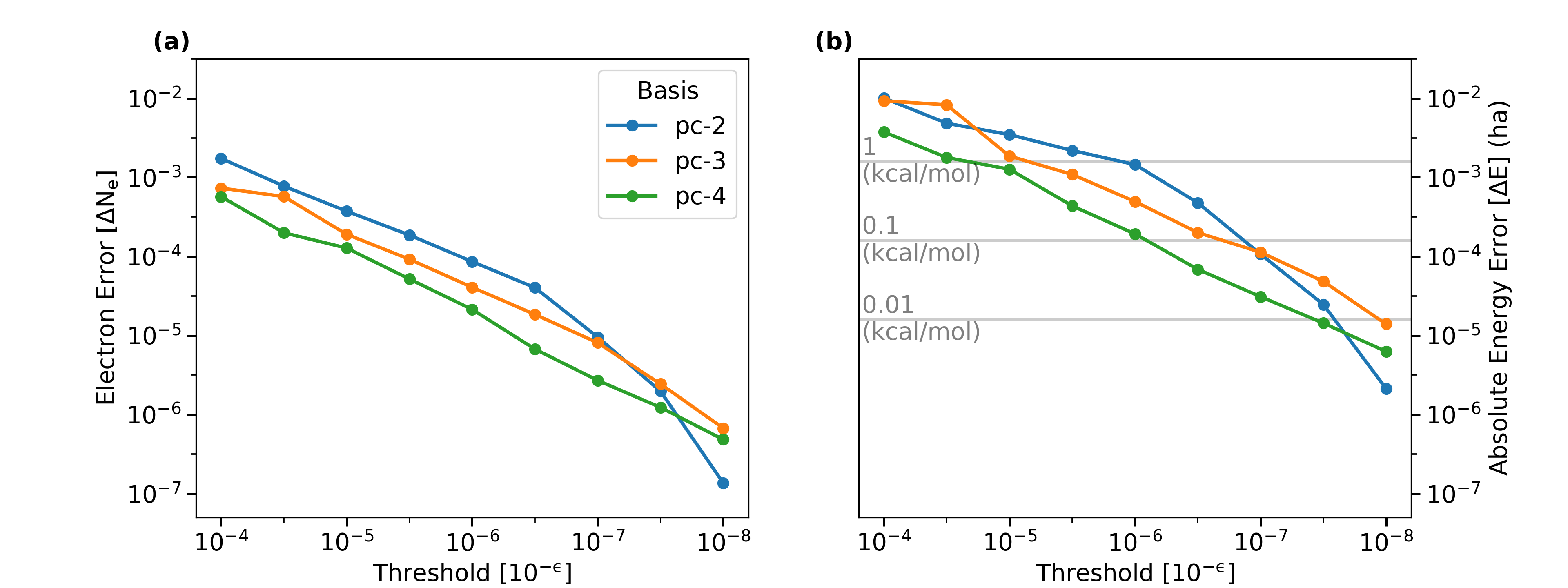}
    \caption{
    Electron and absolute energy error for pc-2, pc-3, and pc-4 basis sets for \ce{C30H62}.
(a) Pre-SCF electron error as a function of threshold. The largest basis set, pc-4, exhibits the smallest error across the range, while the smallest basis, pc-2, generally has the most significant error.
(b) Absolute energy error as a function of threshold. Much like the electron error, error is minimized for pc-4 and maximized for pc-2. 
The electron error and energy error exhibit nearly identical behavior with respect to the threshold.
    }
    \label{fig:30-ane-cardinality-energy}
\end{figure}

To examine the effect of truncating the NAO representation on reproducing the SCF electron density, Figure \ref{fig:30-ane-cardinality-energy}(a) reports the electron error (Eq. \ref{eq:electron-loss}) as a function of threshold across the different basis set cardinalities. For a given threshold, the overall magnitude of the error is roughly the same across basis sets, though it typically decreases slightly with increasing basis size. Across most of the threshold range, the log of the error exhibits an almost perfectly linear dependence on the log of the threshold (i.e., linear with  $\epsilon$ itself) for all basis sets, with very similar slopes between pc-2, pc-3, and pc-4. An exception appears at the tightest thresholds for pc-2, where it is becoming incompressible. These results indicate reasonable $\epsilon$-based control over how well the electron density is represented in the compressed basis.

Figure \ref{fig:30-ane-cardinality-energy}(b) summarizes the absolute energy errors over the same threshold range. It is encouraging to see that the trends for the absolute energy errors closely mirror those seen for electron-count errors; the overall error decreases as basis size increases, and the slope of error versus threshold is again nearly identical across different basis cardinalities. Evidently, there is a strong correlation between the two types of error. Despite this correlation, the energy error displays slightly more variability than the electron error, particularly at the loosest thresholds. The threshold range explored here produces absolute-energy errors in the micro to milli-hartree regime, corresponding to error magnitudes spanning roughly 0.01  to 1 kcal/mol.

Taken together, panels (a) and (b) of Figure \ref{fig:30-ane-cardinality-energy} demonstrate that although basis size affects the magnitude of both electron and energy errors, the functional form of the error threshold relationship is remarkably invariant. Across pc-2, pc-3, and pc-4, the slope and overall shape of the error versus threshold curves are effectively identical, suggesting that the truncation behavior is determined primarily by the spectrum of the system, as shown in Figure \ref{fig:30-ane-cardinality}, rather than by the specific basis cardinality. This consistency indicates that the threshold parameter governs error in a robust and transferable manner, enabling predictable tuning of accuracy independent of the underlying basis size.

\subsection{Errors in relative energies versus threshold}
\begin{figure}[h]
    \centering
    \includegraphics[width=1\linewidth]{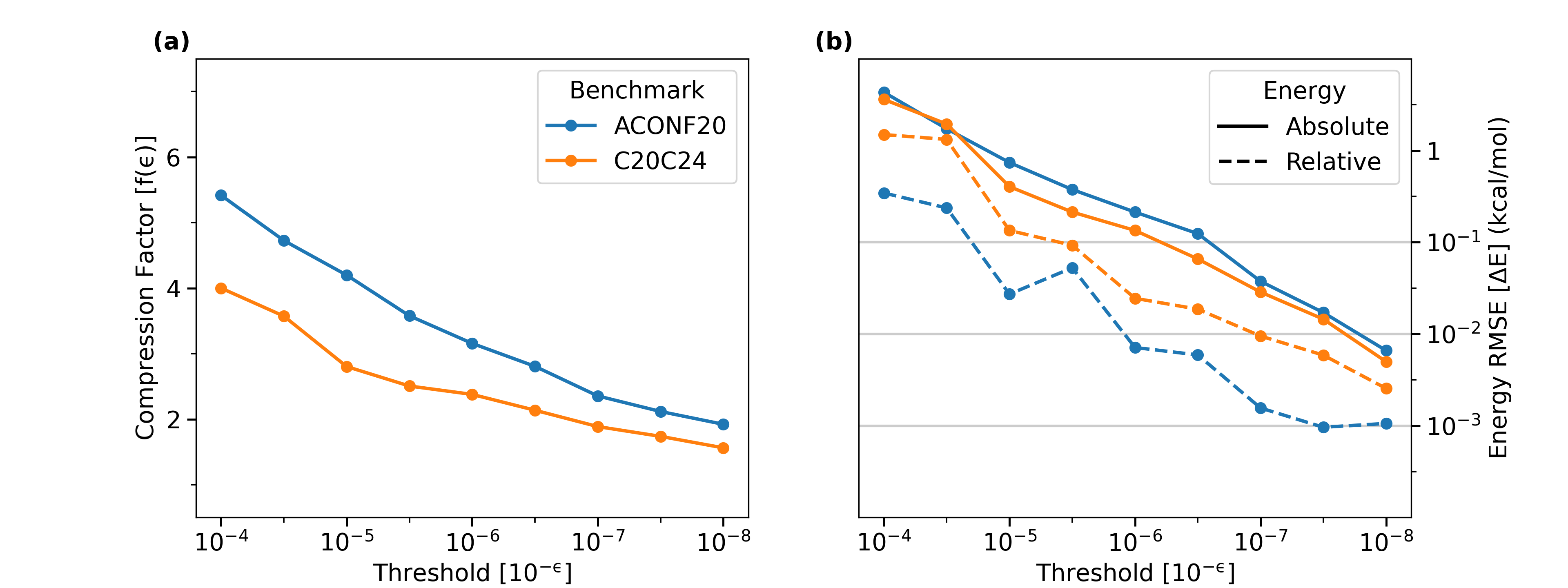}
    \caption{
    Compressibility and energy error for isomerization benchmarks (ACONF20 and C20C24).
    (a) The minimum compression factor of the benchmark versus the threshold. ACONF20 shows significantly higher compressibility across the entire threshold range than C20C24. At worst, both species show 2-fold compressibility and at best 5-fold.
    (b) Absolute and relative energy RMSEs versus threshold. Solid and dashed lines represent the absolute and relative energies RMSE of each benchmark, respectively. ACONF20 relative energy error is an order of magnitude smaller than the absolute energy. C20C24 also has a smaller relative energy error than absolute, but with smaller differences.
    C20C24 exhibits less compressibility and error cancellation than ACONF20.
    }
    \label{fig:benchmark-isomerization}
\end{figure}

Having reported encouraging results for the control of absolute errors in SCF energies in the compressed basis, and direct representation of the full AO density in the compressed basis, we next report our explorations on how errors in relative energies (using the pc-3 AO basis) depend on $\epsilon$. We begin with relative conformational energies for \CH{20}{42} alkane chains (ACONF20)\cite{ehlert2022conformational}, and relative isomer energies for the \ce{C20} and \ce{C24} species,\cite{Manna:2016,liang2025gold} as summarized in Figure \ref{fig:benchmark-isomerization}. Panel (a) shows that the (minimum) compression factor (of the dataset for given $\epsilon$) differs significantly between ACONF20 and C20C24. ACONF20 yields roughly 30\% larger compression factors for a chosen $\epsilon$, consistent with the smaller numbers of $n^\mathrm{th}$ neighbors due to its linear backbone. However, both sets behave qualitatively like pc-3 for all-trans \ce{C30H62} in Figure \ref{fig:30-ane-cardinality}. Figure \ref{fig:benchmark-isomerization}(b) shows energy errors for these two data sets are comparable in absolute magnitude across the threshold range. However, better error cancellation in ACONF20 than C20C24 leads to somewhat smaller relative-energy errors. Presumably, this is because each conformation ACONF20 has the same connectivity, while this is not the case in C20C24. Most importantly, for both datasets, the relative errors are below the absolute errors, indicating beneficial cancellation even in the less favorable case.

\begin{figure}[h]
    \centering
    \includegraphics[width=1\linewidth]{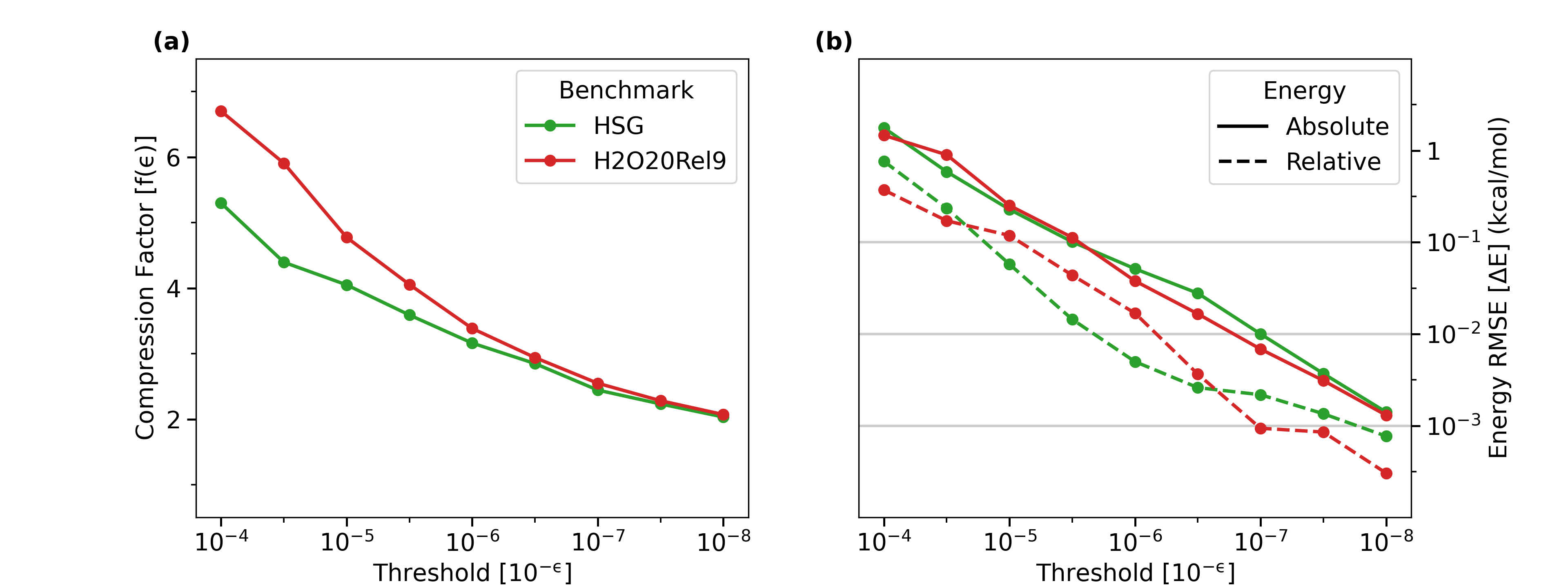}
    \caption{
    Compressibility and energy error for non-covalent interaction benchmarks (HSG and H2O20Rel9).
    (a) The minimum compression factor of the benchmark threshold. H2O20Rel9 exhibits larger compressibility than HSG for loose thresholds.
    (b) Absolute and relative energy RMSEs versus threshold. Solid and dashed lines represent the absolute and relative energies RSME of each benchmark, respectively. Both benchmarks have similar absolute energy errors yet yield vastly different relative energy errors at every threshold. Both show significant decreases in relative energy errors, but HSG shows marked decreases in error between thresholds of $\exp{5}$ to $\exp{7}$.
    With remarkably similar absolute energy errors, the benchmarks differ in compressibility and relative energy errors.
    }
    \label{fig:benchmark-noncovalent}
\end{figure}

Figure \ref{fig:benchmark-noncovalent} presents data (in the same format as Figure \ref{fig:benchmark-isomerization} for two sets of non-covalent interactions: H2O20Rel9\cite{Lao:2015,liang2025gold} and HSG\cite{Marshall:2011,liang2025gold}, specified in Table \ref{tab:benchmark-descriptions}. Panel (a) shows their compressibility profiles, which are similar at tighter thresholds, although H2O20Rel9 shows noticeably greater compression at looser thresholds. 
The fact that bonded connectivity ends at second neighbors in the water molecule makes these clusters (and other molecular clusters, presumably) very compressible, until the highest level of accuracy is required,
a consequence of inherent sparsity in fragmented systems. 
Figure \ref{fig:benchmark-noncovalent}(b) similarly shows that absolute energy errors are nearly identical for the two noncovalent benchmarks. Encouragingly, the relative-energy errors are smaller than the absolute errors; however, the degree of error cancellation varies. 

\begin{figure}[h]
    \centering
    \includegraphics[width=1\linewidth]{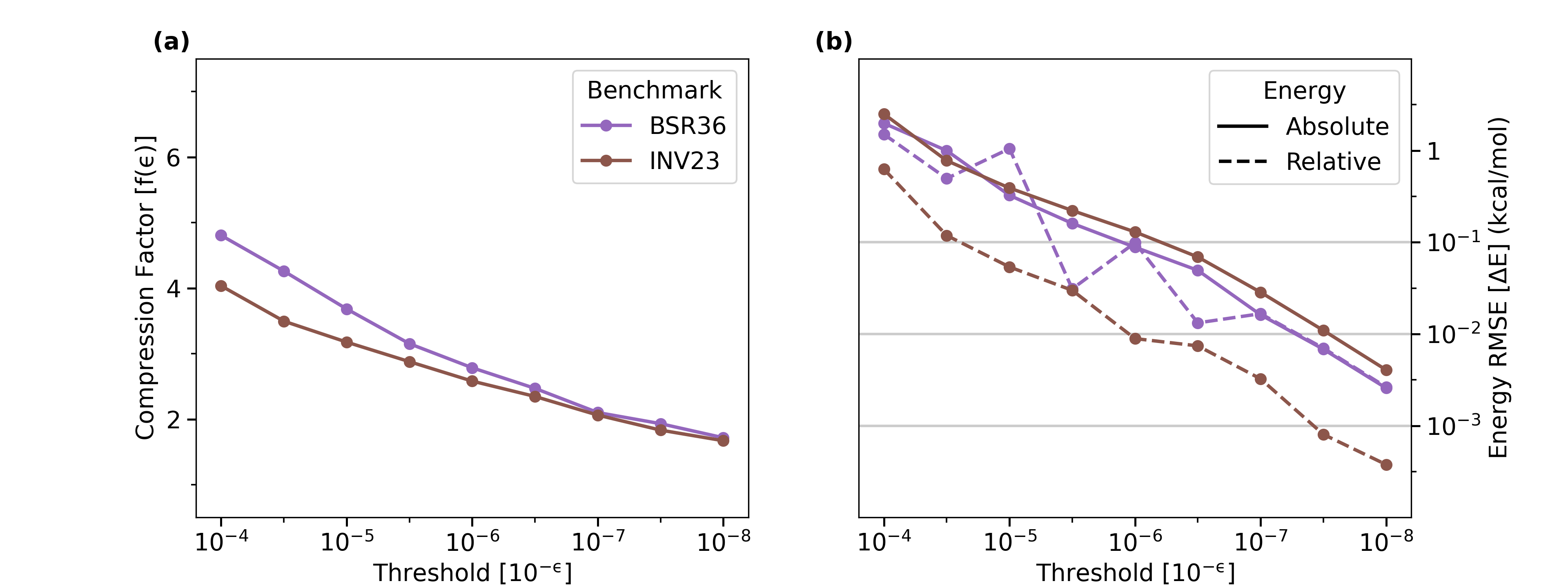}
    \caption{
    Compressibility and energy error for BSR36 and INV23 benchmarks.
    (a) The minimum compression factor of the benchmark versus the threshold. Both benchmarks exhibit similar compressibility except at loose thresholds.
    (b) Absolute and relative energy RMSEs versus threshold. Solid and dashed lines represent the absolute and relative energies RMSE of each benchmark, respectively. Both benchmarks have similar absolute energy errors, yet their relative energy errors differ significantly. The INV23 relative energy is significantly reduced by an order of magnitude compared to its absolute energy. In contrast, the BSR36 relative energy error fluctuates wildly between thresholds.
    With similar compressibility and absolute energy errors, BSR36 exhibits highly variable relative energy, whereas INV23 shows stable relative energy errors.
    }
    \label{fig:benchmark-etc}
\end{figure}

Figure \ref{fig:benchmark-etc} presents data in the same format as above for one thermochemistry data set (BSR36\cite{Goerigk:2017,liang2025gold}) and one barrier height benchmark (INV23\cite{Goerigk:2016,liang2025gold}) (see Table \ref{tab:benchmark-descriptions} for details). Panel (a) shows the behavior of the INV23 compression factor that is quite similar to the C20C24 case shown in Figure \ref{fig:benchmark-isomerization}(a), 
Figure \ref{fig:benchmark-etc}(b) further shows that the absolute energy errors are nearly identical across the two thermochemistry/barrier benchmarks. However, the relative-energy behavior differs substantially between the pair. INV23 exhibits strong error cancellation, yielding relative errors that are well below the absolute errors. By contrast, BSR36 shows substantial variability in relative energies, whereas INV23 exhibits strong error cancellation. It is noteworthy that BSR36 illustrates that relative error can exceed the absolute error. 

\subsection{Error cancellation in relative energies}

The results presented in Figures \ref{fig:benchmark-isomerization}, \ref{fig:benchmark-noncovalent} and \ref{fig:benchmark-etc} showing errors in absolute energies, $\Delta E_i$, and the comparison against errors in relative energies, $\Delta E^{rel}$ invite some further consideration. When is error cancellation in $\Delta E^{rel}$ versus $\Delta E_i$ most favorable and when is it most unfavorable? What are the lessons, if any, for the selection of the threshold, $10^{-\epsilon}$? 

Cases such as ACONF20 and INV23 illustrate highly favorable error cancellation in $\Delta E^{rel}$. We suggest this is because the molecular connectivity associated with the two species involved in the energy difference is very similar. The errors associated with a given choice of $\epsilon$ depend strongly on the number of $n^\mathrm{th}$ neighbors associated with each atom. If this is nearly the same for the species on both sides of the equation determining the energy difference, then the $\epsilon$-related truncation errors can (partially) cancel. Conformational energies of a given alkane backbone as in ACONF20 obey this condition. Some reaction barriers as in INV23 can also come close to meeting this condition. 

Conversely, large changes in connectivity between reactants and products associated with a relative energy will degrade the possibilities for cancellation of the errors in total energies. This is clearly illustrated by the case of the atomization energy of a molecule. The compression for any reasonable $\epsilon$ is exact in the case of the atoms, but of course not exact for any (large) molecule whose atomization energy we seek. Therefore in this extreme example, there is no error cancellation: the error in relative energy is identical with the error in the total energy of the molecule. Therefore a conservative choice of $\epsilon$ should not assume error cancellation in relative energies because it will not always happen.

In some (probably unusual) cases, the errors in relative energies can be \textit{larger} than for total energies, as seen for bond-separation reactions in BSR36 in Figure \ref{fig:benchmark-etc}. The origin of this surprising result is due to some very large stoichiometric coefficients entering the BSR36 relative energies, as shown in Table \ref{tab:BSR36} for select reactions. Large stoichiometric coefficients can greatly amplify the $\epsilon$-based errors for such species.  The main offender is \CH{2}{6}. In contrast, no such amplification occurs for \CH{}{4}, which, like atoms,   exhibits no compression error (e.g. see Figure \ref{fig:n-anes-decay}). While the \CH{}{4} coefficient is commensurate with \CH{2}{6}, the former is exactly compressed and the latter is not, so the value of $\Delta E^{rel}$ is essentially a large multiple of the \ce{C2H6} error. The $\Delta E^{rel}$ is much larger than the absolute error of any input total energy.

\begin{table}[h]
    \centering
    \begin{threeparttable}
    \begin{tabular}{l|lllllll}
        Reaction & $c_\mathrm{{C_{2}H_{6}}}$ & $c_\mathrm{CH_{4}}$ & $c_\mathrm{c_n}$\tnote{1} & $\paran{c\Delta E}_\mathrm{{C_{2}H_{6}}}$ & $\paran{c\Delta E}_\mathrm{CH_{4}}$ & $\paran{c\Delta E}_\mathrm{c_n}$\tnote{1} & $\Delta E^{rel}$ \\
        \hline
        \hline
        BSR36\_32 & 12 & -14 & -1 & 1.84 & 0.0 & -0.39 & 1.45 \\
        BSR36\_34 & 14 & -16 & -1 & 2.14 & 0.0 & -0.51 & 1.63 \\
        BSR36\_35 & 16 & -18 & -1 & 2.45 & 0.0 & -0.48 & 1.97 \\
        BSR36\_36 & 18 & -22 & -1 & 2.75 & 0.0 & -0.55 & 2.20 \\
    \end{tabular}
    \begin{tablenotes}
    \item[1] \textbf{Molecules }$\mathbf{c_n}$\textbf{.} BSR36\_c1, BSR36\_c3, BSR36\_c4, BSR36\_c5
    \end{tablenotes}
    \end{threeparttable}
    \caption{Selected reaction energies ($\Delta E^{rel}$) and respective stoichiometric coefficients ($c$) at a threshold of $\exp{5}$. Large stoichiometric coefficients amplify underlying errors, as shown by the increasing \CH{2}{6} coefficients and the resulting $\Delta E$. The lack of error in \CH{}{4} diminishes possible error cancellation and amplifies error due to the large \CH{2}{6} coefficient.}
    \label{tab:BSR36}
\end{table}

How should such an issue be handled to avoid large errors? One possibility is that the choice of $\epsilon$ for a molecule involved in a relative energy should be connected to its stoichiometric coefficients. We have empirically observed that the relationship between the log of absolute error and the log of threshold (i.e. $\epsilon$) is roughly linear:
\begin{equation}
    \log \Delta E \approx - k \epsilon 
\end{equation}
For long-chain alkanes in the pc-3 basis,  Figure \ref{fig:30-ane-cardinality-energy}(b) suggests $k\sim 1/2$.
One can then estimate the change in $\epsilon$ necessary to decrease the energy error by a factor of 10 if a stoichiometric coefficient of 10 is encountered. For a long-chain alkane, one should increment $\epsilon$ by 2 ($\epsilon+2 \leftarrow \epsilon$) in order to roughly preserve the expected accuracy.

\section{Conclusions}
\label{sec:Conclusions}



In this work, we have proposed and implemented a method for atom-centered compression of large atomic orbital basis sets, yielding highly contracted sets of NAOs that can be truncated by occupation number. Construction of the NAOs is achieved by transforming the SCF density matrix into a one-center orthogonalized representation and diagonalizing its atomic blocks in this representation. The resulting eigenvectors are the NAOs, and the corresponding eigenvalues are their occupation numbers. Due to non-orthogonality effects between atoms, the occupation numbers of fully occupied AOs are not exactly unity, although they remain relatively close for core orbitals, regardless of the size of the underlying basis. The following levels are most strongly occupied and resemble molecule-adapted valence atomic orbitals. 
Together, these NAOs define a minimal atomic orbital basis that does not fully span the occupied space. 

The spectrum of NAO occupation numbers typically exhibits a pronounced gap separating the effective minimal basis from the remaining NAOs, which have much smaller occupation numbers. For large molecules (hydrocarbon chains were tested), the remaining small occupation numbers exhibit approximately exponential decay with their index. 
This observed behavior is the essential observation that suggests it may be useful to employ a truncated set of NAOs based on a threshold to closely reproduce large basis SCF energies using a much smaller set of NAOs.  

We define the compression ratio, $f(\epsilon)$ as the ratio of the number of AOs, $N$, to the number of retained NAOs, $M(\epsilon)$. 
Larger compression ratios $f(\epsilon)$ for given $\epsilon$ are obtained in larger basis sets. Small molecules are most highly compressible, and the extent of compressibility appears to approach limiting values that depend on the number of $n^{\mathrm{th}}$ neighbors for interior atoms, based on bonded connectivity.

The SCF energy is evaluated in the compressed representation to assess the energetic consequences of the occupation number threshold. We presented extensive numerical results demonstrating that truncation offers a controllable route for considerably compressing the size of large AO basis sets (e.g., pc-2, pc-3, and pc-4) while maintaining high accuracy as measured by absolute and relative energies across non-covalent energy differences, conformation energies, isomerization energies, thermochemistry, and barrier heights. In particular, errors in relative energies were typically much smaller than the corresponding absolute errors.

Taken together, these findings establish a proof of concept that AO-basis compression using NAOs with a minimum occupation-number threshold provides a viable and controllable approximation to SCF energies in extensive basis sets. These results pave the way for future work focused on leveraging compressed NAOs to accelerate SCF calculations in these large-basis regimes. We will report further developments in due course, addressing two key issues.

The first issue concerns how to perform the compression without requiring a nearly converged density matrix in the large basis. Dual-basis ideas \cite{liang2004approaching,mao2016approaching} appear to be a promising route: they require only a converged density in a smaller basis and a single Fock-matrix evaluation in the large basis. This one-step corrected density matrix can then be used to generate accurate compressed NAOs in the large basis set. 

The second issue is how to exploit the compressed basis to accelerate the SCF procedure efficiently. Our pilot implementation already carries out all linear algebra in the compressed basis, reducing the asymptotically rate-determining steps by a factor of $(N/M)^3$. In standard SCF, the dominant computational cost still arises from the 4-center 2-electron integrals, despite their quadratic scaling, for all but extremely large molecules. Therefore, accelerating the diagonalization step alone yields little benefit. In contrast, for resolution-of-the-identity (RI)–based methods \cite{weigend2002fully,Manzer:2015b}, the most time-consuming step becomes the linear algebra, even for medium-sized molecules, due to its high scaling. In this case, compression is a desirable strategy, with potential speedups in the linear-algebra component ranging from $\mathcal{O}[(N/M)^2]$ to at most $\mathcal{O}[(N/M)^4]$, depending on how many AO indices are incorporated into the compression scheme.


\section*{Conflicts of Interest}
MHG is a part-owner of Q-Chem Inc., which is the software platform used to implement the algorithms described here.

\section*{Acknowledgments}
This research was supported by the Gas Phase Chemical Physics Program in the Chemical Sciences, Geosciences, and Bio-sciences Division of the Office of Basic Energy Sciences of the U.S. Department of Energy under Contract No. DE-AC02-05CH11231.

\section*{Supporting Information}

The following raw data files are provided:
\begin{itemize}
\item \textit{occupation-numbers.csv} — occupation numbers for the 
\nane, \CH{30}{32}, and \CH{30}{2} systems.
\item \textit{absolute-energies.csv} — absolute energies for the benchmarked systems presented in Table \ref{tab:benchmark-descriptions}.
\item \textit{stoichiometric.csv} — stoichiometric coefficients used in computing the relative-energy benchmarks.
\end{itemize}

\providecommand{\latin}[1]{#1}
\makeatletter
\providecommand{\doi}
  {\begingroup\let\do\@makeother\dospecials
  \catcode`\{=1 \catcode`\}=2 \doi@aux}
\providecommand{\doi@aux}[1]{\endgroup\texttt{#1}}
\makeatother
\providecommand*\mcitethebibliography{\thebibliography}
\csname @ifundefined\endcsname{endmcitethebibliography}  {\let\endmcitethebibliography\endthebibliography}{}

\newpage
\begin{center}
\centering $\mathbf{TOC~Graphic}$
\end{center}
\begin{figure}[h]
    \centering
   \includegraphics[width=9cm]{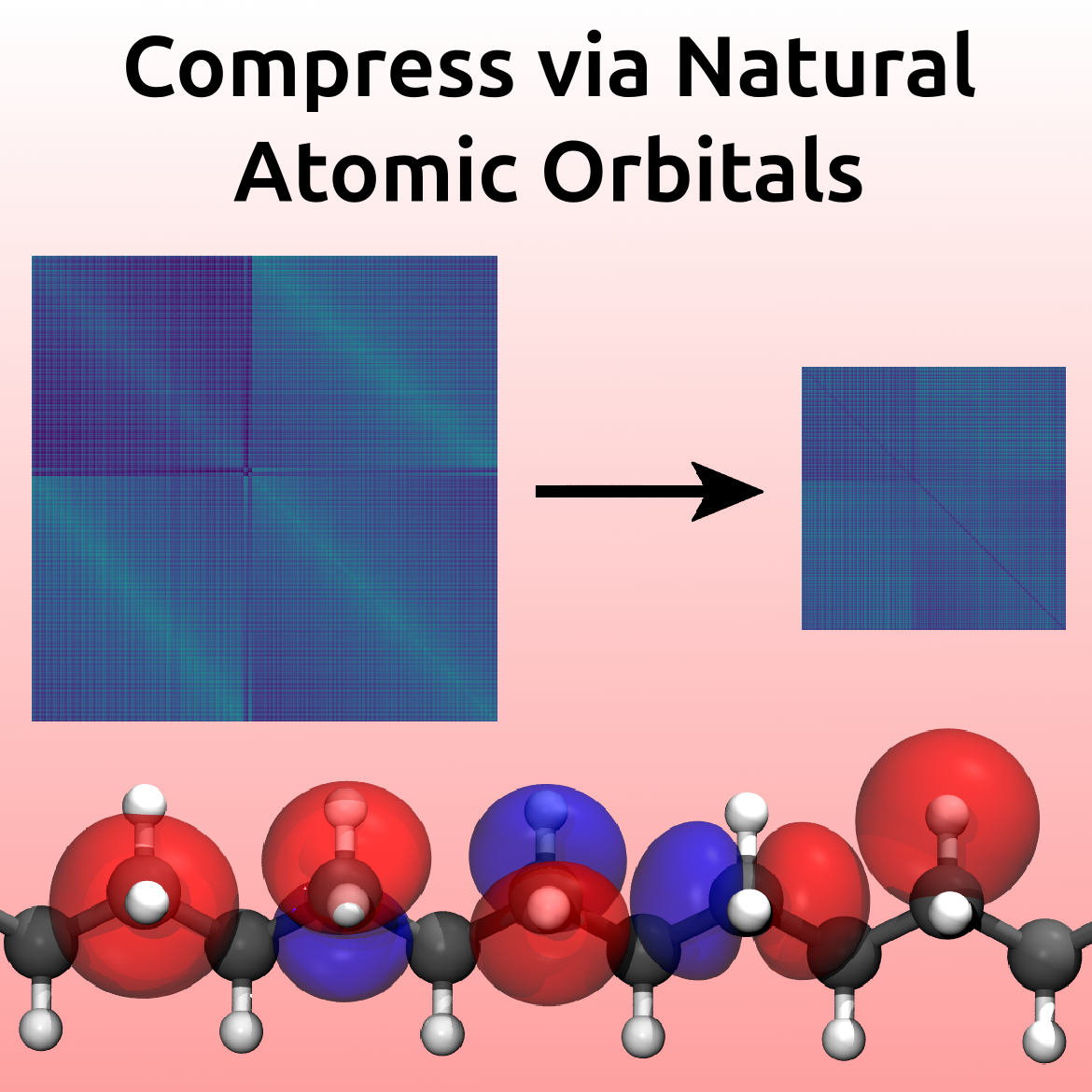}
\end{figure}

\end{document}